\documentstyle[12pt,epsf]{article}

\catcode`\@=11
\long\def\@makefntext#1{
\protect\noindent \hbox to 3.2pt {\hskip-.9pt
$^{{\ninerm\@thefnmark}}$\hfil}#1\hfill}                

\def\@makefnmark{\hbox to 0pt{$^{\@thefnmark}$\hss}}  

\def\ps@myheadings{\let\@mkboth\@gobbletwo
\def\@oddhead{\hbox{}
\rightmark\hfil\ninerm\thepage}
\def\@oddfoot{}\def\@evenhead{\ninerm\thepage\hfil
\leftmark\hbox{}}\def\@evenfoot{}
\def\sectionmark##1{}\def\subsectionmark##1{}}

\renewcommand{\thefootnote}{\fnsymbol{footnote}}

\def\sectionc{\@startsection {section}{1}{\z@}{-3.5ex plus -1ex minus
    -.2ex}{2.3ex plus .2ex}{\bf }}
\def\subsectionc{\@startsection{subsection}{2}{\z@}{-3.25ex plus -1ex minus
   -.2ex}{1.5ex plus .2ex}{\it }}
\renewcommand{\section}[1]{\sectionc{#1}\hspace*{\parindent}}
\renewcommand{\subsection}[1]{\subsectionc{#1}\hspace*{\parindent}}
\newcounter{appendixc}
\newcounter{subappendixc}[appendixc]
\newcounter{subsubappendixc}[subappendixc]

\renewcommand{\appendix}[1] {\vspace*{0.6cm}
        \refstepcounter{appendixc}
        \setcounter{figure}{0}
        \setcounter{table}{0}
        \setcounter{equation}{0}
        \renewcommand{\thefigure}{\Alph{appendixc}.\arabic{figure}}
        \renewcommand{\thetable}{\Alph{appendixc}.\arabic{table}}
        \renewcommand{\theappendixc}{\Alph{appendixc}}
        \renewcommand{\theequation}{\Alph{appendixc}.\arabic{equation}}
        \noindent{\bf Appendix \theappendixc #1}\par\vspace*{0.4cm}}

\def\abstracts#1{{
        \centering{\begin{minipage}{13.2truecm}\footnotesize\baselineskip=13pt\noindent
        \parindent=0pt #1
        \end{minipage}}\par}}

\renewenvironment{thebibliography}[1]
        {\begin{list}{\arabic{enumi}.}
        {\usecounter{enumi}\setlength{\parsep}{0pt}
\setlength{\leftmargin 0.75cm}{\rightmargin 0pt}
         \setlength{\itemsep}{0pt} \settowidth
        {\labelwidth}{#1.}\sloppy}}{\end{list}}

\newcounter{itemlistc}
\newcounter{romanlistc}
\newcounter{alphlistc}
\newcounter{arabiclistc}

\newcommand{\fcaption}[1]{
        \refstepcounter{figure}
        \setbox\@tempboxa = \hbox{\footnotesize Figure~\thefigure. #1}
        \ifdim \wd\@tempboxa > 6in
           {\begin{center}
        \parbox{6in}{\footnotesize\baselineskip=13pt Figure~\thefigure. #1}
            \end{center}}
        \else
             {\begin{center}
             {\footnotesize Figure~\thefigure. #1}
              \end{center}}
        \fi}

\newcommand{\tcaption}[1]{
        \refstepcounter{table}
        \setbox\@tempboxa = \hbox{\footnotesize Table~\thetable. #1}
        \ifdim \wd\@tempboxa > 6in
           {\begin{center}
        \parbox{6in}{\footnotesize\baselineskip=13pt Table~\thetable. #1}
            \end{center}}
        \else
             {\begin{center}
             {\footnotesize Table~\thetable. #1}
              \end{center}}
        \fi}

\def\@citex[#1]#2{\if@filesw\immediate\write\@auxout
        {\string\citation{#2}}\fi
\def\@citea{}\@cite{\@for\@citeb:=#2\do
        {\@citea\def\@citea{,}\@ifundefined
        {b@\@citeb}{{\bf ?}\@warning
        {Citation `\@citeb' on page \thepage \space undefined}}
        {\csname b@\@citeb\endcsname}}}{#1}}

\newif\if@cghi
\def\cite{\@cghitrue\@ifnextchar [{\@tempswatrue
        \@citex}{\@tempswafalse\@citex[]}}
\def\citelow{\@cghifalse\@ifnextchar [{\@tempswatrue
        \@citex}{\@tempswafalse\@citex[]}}
\def\@cite#1#2{{$\null^{#1}$\if@tempswa\typeout
        {IJCGA warning: optional citation argument
        ignored: `#2'} \fi}}

 1
 1
 1

\font\ninerm=cmr9

\newcommand{\beq}{\begin{eqnarray}}
\newcommand{\eeq}{\end{eqnarray}}

\newcommand{\cket}[1]{| #1 \rangle}
\newcommand{\bra}[1]{\langle #1 |}
\let \ket=\cket

\newcommand{\pslash}{{p\hspace{-5pt}/}}

\newcommand{\gpNN}{g_{\pi NN}}
\newcommand{\geNN}{g_{\eta NN}}
\newcommand{\gpNR}{g_{\pi NN^*}}
\newcommand{\geNR}{g_{\eta NN^*}}
\newcommand{\vev}[1]{\langle #1 \rangle}
\newcommand{\GeV}{\: {\rm GeV}}
\begin{document}

\centerline{\normalsize\bf Negative Parity Baryons in the QCD Sum Rule}
\baselineskip=15pt

\vspace*{0.6cm}
\centerline{\footnotesize 
M.~Oka\footnote{e-mail address: oka@th.phys.titech.ac.jp}{}
\ and D.~Jido}

\baselineskip=13pt
\centerline{\footnotesize\it Department of Physics, 
Tokyo Institute of Technology}
\centerline{\footnotesize\it Meguro, Tokyo 152, Japan}

\vspace*{0.3cm}

\centerline{\footnotesize A.~Hosaka}

\centerline{\footnotesize\it Numazu College of Technology}
\centerline{\footnotesize\it 3600 Ooka, Numazu 410 Japan}

\baselineskip=13pt

\vspace*{0.3cm}
\vspace*{0.3cm}

\vspace*{0.6cm}
\abstracts{Masses and couplings of the negative parity excited baryons are 
studied in the QCD sum rule.  Separation of the negative-parity spectrum 
is proposed and is applied to the flavor octet and singlet baryons.  We find 
that the quark condensate is responsible for the mass splitting of the ground 
and the negative-parity excited states.  This is expected from the chiral 
symmetry and supports the idea that the negative-parity baryon forms
a parity doublet with the ground state.  
The meson-baryon coupling constants are 
also computed for the excited states in the QCD sum rule.   It is found that 
the $\pi NN^*$ coupling vanishes in the chiral limit.}

\normalsize\baselineskip=15pt
\setcounter{footnote}{0}
\renewcommand{\thefootnote}{\alph{footnote}}

\section{Introduction} 
The negative parity baryons have been successfully described by 
the (nonrelativistic) quark model as one-quark excited states 
belonging to the SU(6) $\underline{70}$ representation\cite{ik}. 
The observed spectrum tends to agree with the prediction, 
although some of the states, such as $\Lambda(1405)$, $N(1535)$, 
have irregular masses and non-natural 
decay rates. Many refinements were proposed to achieve a 
quantitative agreement.

Yet, our understanding of the hadron physics insists that the role of 
the chiral symmetry must be important even in the baryon states. 
Indeed, the chiral symmetry suggests that the positive and negative 
parity states are paired into a parity doublet and the pair would be 
degenerate when the chiral symmetry is restored. Because the 
nonrelativistic quark model does not observe the chiral symmetry, 
a different approach is anticipated to understand the 
chiral structure of baryons, both the positive and negative parity 
states.

The technique of the QCD sum rule relates the hadron properties to the 
QCD parameters and is a powerful tool to extract the hadron properties 
from QCD, the first principle of the strong interaction\cite{svz,rry}.  
The QCD sum rule for the baryon, first proposed by Ioffe\cite{i},
considers a correlation function of an interpolating field (IF) that couples 
to the baryon state in question.  
The correlation function is computed in two ways, 
phenomenologically and theoretically, and their equality gives a sum rule.
The nonperturbative effects, 
such as the quark condensate $\vev{\bar{q}q}$ and $\vev{{\alpha_{s} 
\over \pi} GG}$, are included as the power corrections of the operator
product expansion of the correlator in the 
theoretical side.  The quark condensate $\vev{\bar{q}q}$, which is the 
order parameter of the chiral symmetry breaking, gives effects of the 
chiral symmetry breaking to the hadron spectrum in the QCD sum rule.

We here study the masses and couplings of the negative-parity 
baryons in the QCD sum rule approach\cite{jko,jo,joh}.  
Our aim is to study roles of the 
chiral symmetry in baryon excitations.  We first present
a technique to separate the positive and
negative parity states in the sum rule,  and then calculate 
their masses\cite{jko}.  It is important to separate 
contribution of the negative-parity baryon ($B_{-}$) from that of 
the positive-parity baryon ($B_{+}$), since the IF for the baryon couples 
to both  $B_{-}$ and $B_{+}$ simultaneously\cite{i,cdks}.
In order to separate the $B_{-}$ 
contribution we use the ``old-fashioned'' correlation function defined 
as
\begin{equation}
   \Pi(p) = i \int d^{4} x e^{i p \cdot x} \theta(x_{0}) \bra{0} J_{B}(x) 
   \bar{J}_{B}(0) \cket{0}, \label{eq:oldfco}
\end{equation}
and construct sum rules in the rest 
frame ($\vec{p}=0$).  Our approach is suitable for investigating
the mass splitting, because $B_{+}$ and $B_{-}$ can be treated 
simultaneously in this sum rule.
We apply the technique
not only to the flavor octet baryons (nucleons and hyperons) but also 
to the flavor singlet baryon $\Lambda_{S}$. 

\section{Negative-parity Baryons in the QCD Sum Rule}
It is important to note that the interpolation field (IF) for a baryon 
does not specify its parity. In the mesonic case, the parity of the state 
is directly connected to the parity of the IF.
For instance, the IF 
for the $\rho^{+}$-meson is a vector current, $\bar{d}\gamma_{\mu} u$, 
while that 
for $a_{1}^{+}$-meson, which is the chiral partner of $\rho$, is 
an axial vector current, $\bar{d}\gamma_{\mu}\gamma_{5}u$.  
This is not the case for the baryon.
The IF for the nucleon, for instance, is given by
\begin{equation}
   J_{N}(x) = J_{+}(x) = \varepsilon_{abc} [(u_{a}(x)Cd_{b}(x)) 
      \gamma_{5} u_{c}(x) + t (u_{a}(x) C \gamma_{5} d_{b}(x)) 
u_{c}(x)],
  \label{eq:nucur} 
\end{equation}
where $a$, $b$ and $c$ are color indices, $C = i \gamma_{2} 
\gamma_{0}$ (standard notation) is for the charge conjugation and $t$ 
is a real parameter representing the mixing of two independent IFs. 
Then it may seem that the negative-parity baryon couples to the IF 
$J_{-} \equiv i\gamma_{5} J_{+}$ because 
multiplying $i \gamma_{5}$ changes the ``parity''.
If one supposes that the correlation function of $J_{N}$ is given by 
\begin{equation}
    \Pi_{+}(p) = p_{\mu}\gamma^{\mu}\Pi_{1}(p^{2}) + \Pi_{2}(p^{2}),
\end{equation}
then the correlation function for $J_{-}$ can be written as 
\begin{equation}
    \Pi_{-}(p) = -\gamma_{5} \Pi_{+}(p)\gamma_{5} = p_{\mu} 
    \gamma^{\mu}\Pi_{1}(p^{2}) - \Pi_{2}(p^{2}).  
\end{equation}
The difference between $\Pi_{+}$ and $\Pi_{-}$ appears only in the sign in 
front of $\Pi_{2}(p^{2})$.  That is, they are given by the same functions 
$\Pi_{1}(p^{2})$ and $\Pi_{2}(p^{2})$.  
This means that the information of the negative-parity nucleon is already 
included in $\Pi_{+}(p)$ since $J_{N}$ couples not only to the positive-parity 
nucleon but also to the negative-parity excited state\cite{cdks}.  It is 
easy to see this from
\begin{equation}
    \bra{0}J_{+}\cket{B^{-}} \bra{B^{-}}\bar{J_{+}}\cket{0} = 
    - \gamma_{5}\bra{0}J_{-}\cket{B^{-}}\bra{B^{-}} 
    \bar{J_{-}}\cket{0} \gamma_{5},
\end{equation}
where $\cket{B^{-}}$ denotes a single baryon state with negative parity.  
$J_{-}$ couples to the positive-parity states in the same way.

 We have proposed a formulation for separating the negative-parity 
contribution from the sum rule\cite{jko}.
To do so, we use the ``old-fashioned'' correlation 
function (\ref{eq:oldfco}).  In the zero-width resonance 
approximation, we write the imaginary part in the rest frame $\vec{p} 
= 0$ as
\begin{eqnarray}
  {\rm Im} \, \Pi(p_{0}) & = & \sum_{n} \left[
    (\lambda_{n}^{+})^{2} {\gamma_{0} + 1\over 2} \delta(p_{0} - m_{n}^{+}) + 
    (\lambda_{n}^{-})^{2} {\gamma_{0} - 1\over 2} \delta(p_{0} - m_{n}^{-}) \right] 
  \label{eq:imphe} \\
  & \equiv & \gamma_{0} A(p_{0}) + B(p_{0}) \nonumber,
\end{eqnarray}
where $m_{n}^{\pm}$ denotes the mass of the $n$-th resonance and 
$\lambda_{n}^{\pm}$  the coupling strength of the IF to the 
resonance. 
Now, one sees that $A(p_{0})$ and $B(p_{0})$ are given by
\begin{eqnarray*}
  A(p_{0}) & = & {1 \over 2} \sum_{n}[ (\lambda_{n}^{+})^{2} 
  \delta(p_{0} - m_{n}^{+})  + (\lambda_{n}^{-})^{2} \delta(p_{0} - 
  m_{n}^{-})], \\
  B(p_{0}) & = & {1 \over 2} \sum_{n}[ (\lambda_{n}^{+})^{2} 
  \delta(p_{0} - m_{n}^{+}) - (\lambda_{n}^{-})^{2} \delta(p_{0} - 
  m_{n}^{-})] .
\end{eqnarray*}
and  that the combination $A(p_{0}) + B(p_{0})$ ($A(p_{0}) - 
B(p_{0})$) contains only the positive-parity 
(negative-parity) states.

We, however, can no longer construct sum rules in $p^{2}$-space, since 
the ``old-fashioned'' correlation function is not analytic in $p^{2}$ 
space.  Instead a dispersion relation can be written in the complex $p_{0}$ 
plane, because the correlation function (\ref{eq:oldfco}) is analytic 
in the upper-half region of the complex $p_{0}$ plane.  The 
theoretical side is given by the operator product expansion, which is 
valid at high energy i.e.  $\Pi^{\rm OPE}(p_{0}=Q) \simeq \Pi^{\rm 
Phe}(p_{0}=Q)$ at large $|Q|$.  Using the analyticity we obtain 
\begin{eqnarray}
    \int_{0}^{Q}[A^{\rm OPE}(p_{0}) - A^{\rm Phe}(p_{0}) ] W(p_{0}) \, dp_{0}
        & = & 0, \label{eq:sra} \\
    \int_{0}^{Q}[B^{\rm OPE}(p_{0}) - B^{\rm Phe}(p_{0}) ] W(p_{0}) \, dp_{0} 
        & = & 0,  \label{eq:srb}
\end{eqnarray}
where $W(p_{0})$ is an arbitrary analytic function which is real on 
the real axis.  Note that we use the fact that the imaginary part of the 
correlation vanishes for negative $p_{0}$.

The standard choice of $W$ is the Borel weight $W(p_{0}) = \exp( - \frac{{p_{0}}^{2}}{M^{2}})$.
We take the lowest mass pole and approximate others as 
continuum whose behavior above a threshold $s_{0}^{\pm}$ coincides with
the theoretical side. Then we obtain two sum rules
\begin{eqnarray}
     {1 \over 2} [\tilde{A}^{\rm OPE}(M,s_{0}^{+}) + \tilde{B}^{\rm 
     OPE}(M,s_{0}^{+})] & = & 
     (\lambda^{+})^{2} \exp[{- \frac{(m^{+})^{2}}{M^{2}}}],  \label{eq:srp} \\
     {1 \over 2} [\tilde{A}^{\rm OPE}(M,s_{0}^{-}) - \tilde{B}^{\rm 
     OPE}(M,s_{0}^{-})] & = & 
     (\lambda^{-})^{2} \exp[{- \frac{(m^{-})^{2}}{M^{2}}}], \label{eq:srn}
\end{eqnarray} 
where 
\begin{eqnarray*}
     \tilde{A}^{\rm OPE}(M,s_{0}^{+}) = \int^{s_{0}^{+}}_{0}dp_{0}
     A^{\rm OPE}(p_{0}) \exp(- \frac{p^{2}_{0}}{M^{2}}) , \\
     \tilde{B}^{\rm OPE}(M,s_{0}^{-}) = \int^{s_{0}^{-}}_{0}dp_{0}
     B^{\rm OPE}(p_{0}) \exp(- \frac{p^{2}_{0}}{M^{2}}).
\end{eqnarray*}
The first sum rule is for the baryons with positive parity and the 
second one is for negative-parity baryons.  In these sum rules, we 
allow the threshold to be different for each parity.


\section{Borel Sum Rules for Baryon Masses}

It is important to choose an appropriate interpolating field (IF)
in the QCD sum rule\cite{jo}.  The IF
should have the same quantum numbers as the baryon in question, so
that it creates or annihilates a single particle state of the
baryon from the vacuum. For the spin ${1 \over 2}$ 
octet baryon two independent IFs can be constructed
without a derivative\cite{ept}. The IF for the nucleon is given above,
eq.(\ref{eq:nucur}).
If we choose the parameter $t=-1$ and use the Fierz transformation, 
it is reduced to the Ioffe's IF\cite{i}.  This choice seems adjusted to the 
positive-parity baryon state.  Instead, we found that $t=0.8$ is 
appropriate for the negative-parity resonance\cite{jko}.  
The IFs for the $\Sigma$ and $\Xi$ baryons are obtained by 
replacing a $d$-quark by an $s$-quark or $u$ by $s$ in eq.(\ref{eq:nucur}) ,
\begin{equation}
   J_{\Sigma^{+}}(x) = \varepsilon_{abc} [(u_{a}(x)Cs_{b}(x)) 
      \gamma_{5} u_{c}(x) + t (u_{a}(x) C \gamma_{5} s_{b}(x)) 
u_{c}(x)].
  \label{eq:sicur} 
\end{equation}
\begin{equation}
   J_{\Xi^{-}}(x) = \varepsilon_{abc} [(s_{a}(x)Cd_{b}(x)) 
      \gamma_{5} s_{c}(x) + t (s_{a}(x) C \gamma_{5} d_{b}(x)) 
s_{c}(x)].
  \label{eq:xicur} 
\end{equation}
The IF for $\Lambda$ is given by 
\begin{eqnarray}
  J_{\Lambda}(x) & = & \varepsilon_{abc} [(d_{a}(x)Cs_{b}(x)) 
      \gamma_{5} u_{c}(x) + (s_{a}(x)Cu_{b}(x)) \gamma_{5} d_{c}(x)  
      \nonumber \\
  & &   - 2 (u_{a}(x)Cd_{b}(x)) \gamma_{5} s_{c}(x) 
\label{eq:lacur} \\
  & & + t \{ (d_{a}(x)C\gamma_{5} s_{b}(x)) u_{c}(x) + 
       (s_{a}(x)C\gamma_{5} u_{b}(x)) d_{c}(x)  \nonumber \\
  & &  - 2 (u_{a}(x)C\gamma_{5}d_{b}(x)) s_{c}(x) \}] \nonumber  
\end{eqnarray}
The parameter $t=-0.8$ is used also for the $\Sigma^*$, $\Lambda^*$ and
$\Xi^*$ resonances.

The IF for the flavor singlet baryon $\Lambda_S$ is uniquely given by 
the flavor antisymmetric combination of the quark operators:
\begin{eqnarray}
  J_{\Lambda_S}(x) & = & \varepsilon_{abc} 
	[(u_{a}(x)C\gamma_{5}d_{b}(x))s_{c}(x)
     - (u_{a}(x) C d_{b}(x)) \gamma_{5} s_{c}(x) \nonumber \\
  & &  - (u_{a}(x)C\gamma_{5}\gamma_{\mu} d_{b}(x)) \gamma^{\mu} s_{c}(x) ] .
  \label{eq:singcur} 
\end{eqnarray}

The theoretical (OPE) sides  (up to dimension 6)
of the sum rules (\ref{eq:sra}) and (\ref{eq:srb}) for the nucleon 
are given by
\begin{eqnarray}
    {\rm Im} A^{\rm OPE}(p_{0}) & = & {5 + 2 t + 5 t^2 \over 2^{10} 
    \pi^{4}} p_{0}^{\,5} 
    \theta (p_{0})
    + { 5 + 2 t + 5 t^{2} \over 2^{9} \pi^{2}} p_{0} \theta (p_{0}) \langle 
    {\alpha_{s} \over \pi} GG \rangle  \nonumber \\
    & -& {5 + 2 t - 7 t^{2}  \over 12 } \delta(p_{0})     
    \langle  \bar{q} q \rangle ^{2}, \label{eq:nucor} \\
    {\rm Im} B^{\rm OPE}(p_{0}) & =  & 
          -  {7 t^{2}  - 2t -5 \over 32 \pi^{2}} p_{0}^{\,2} 
          \theta(p_{0}) \langle \bar{q} q \rangle -
     {3(1-t^{2}) \over 32 \pi^{2}} \theta(p_{0}) \langle \bar{q} g
    \sigma \cdot G q \rangle . \nonumber
\end{eqnarray}
In these expressions we neglect the up and down quark masses.  

Note that the difference of the sum rules for $B_+$ and $B_-$ is the 
chiral odd term $B(p_{0})$, which is proportional either to the quark 
condensate $\vev{\bar{q}q}$ or to the mixed condensate $\vev{ \bar{q}g 
\sigma \cdot G q}$.  If the chiral symmetry is restored, for instance, at high 
temperature, the $B(p_{0})$ term goes to zero (in the 
chiral limit).  Then the sum rules (\ref{eq:srp}) and (\ref{eq:srn}) 
are identical, and will predict the same masses for the positive and 
negative parity baryons.  This situation is similar to that in the 
linear sigma model for parity-doublet baryons proposed by DeTar and 
Kunihiro\cite{dk}.  There the positive and negative parity baryons
are assumed to form a parity doublet and the Lagrangian has a chiral 
invariant mass term.  Under the restoration of the chiral symmetry, 
$B_+$ and $B_-$  have the same mass, while in the spontaneous symmetry 
broken phase the mass splitting is proportional to the non vanishing 
vacuum expectation value of sigma.  

\begin{figure}[tb]
\epsfxsize=13cm
\epsfbox{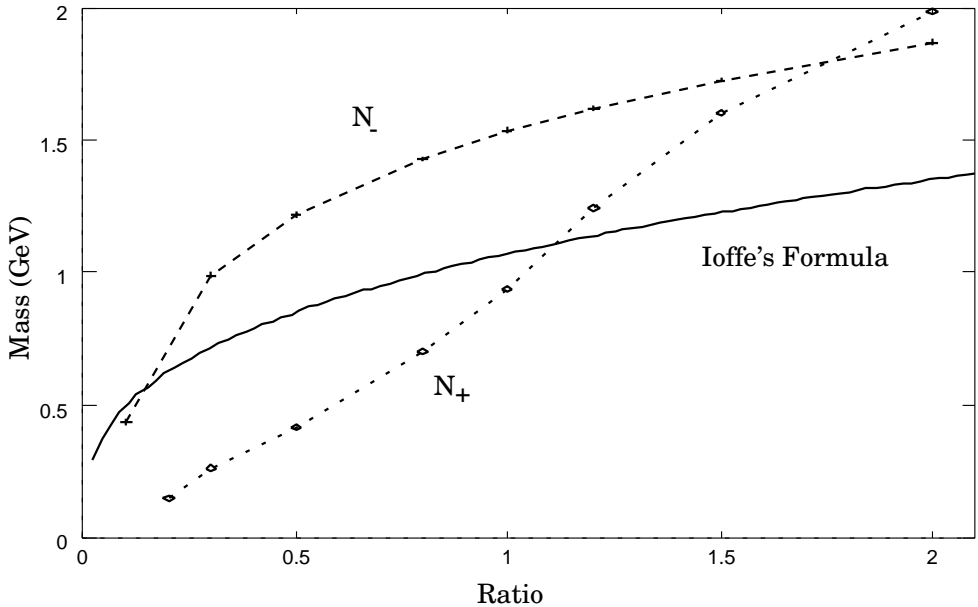}
	 \caption[]{\small Masses of $N^{+}$ and $N^{-}$ at $M=2.5$ GeV 
	 for various values of the quark condensate.  $R$ is the ratio of 
	 $\langle \bar{q} q \rangle $ to its standard value $\langle 
	 \bar{q} q \rangle _{0}$.  The solid line is the Ioffe's formula 
	 (\ref{eq:ioffo}).  }
    \label{fg:nvarq}
\end{figure}

In order to see the effect of the chiral symmetry breaking, we vary 
$\langle \bar{q} q \rangle $ and study its effects.  $\langle \bar{q} 
\sigma \cdot G q \rangle$ is assumed to be proportional to $\langle 
\bar{q} q \rangle$ and therefore is varied together with $\langle 
\bar{q} q \rangle$. 
Fig.\ref{fg:nvarq} shows the masses of $N^+$ and $N^-$ for various values
of $R=\langle \bar{q} q \rangle / \langle \bar{q} q \rangle_0$.
One sees that both the masses of 
$N^{+}$ and $N^{-}$ tend to decrease for $R\to 0$ and become 
degenerate in the limit, although the $R$ 
dependencies are different.  It should be noted that the 
behavior of the $N^{+}$ mass is different from the Ioffe's 
formula\cite{i}
\begin{equation}
m^{+} = [-2 (2\pi)^{2} \langle \bar{q}q \rangle]^{1/3} \label{eq:ioffo}
\end{equation}

We have three phenomenological parameters, the mass $m_{B_{\pm}}$, 
the threshold $s_\pm$ and the coupling strength $\lambda_\pm$ to be 
determined from the sum rules (\ref{eq:srp}) and (\ref{eq:srn}).
These parameters are determined by solving the system of three equations, 
eq. (\ref{eq:srp}) or (\ref{eq:srn}), and its first and second derivatives
with respect to the Borel mass.

The theoretical side depends on the QCD parameters, such as the quark 
mass and the gauge coupling constant, and also on the other parameters 
that describe the properties of the nonperturbative vacuum of QCD, 
such as the quark and gluon condensates.  We take the chiral limit for 
the up and down quarks, i.e.\ $m_{q}=0$, where we use the symbol $q$ 
for the up and down quarks.
We introduce the strange quark mass $m_{s}$, $\chi \equiv 
\vev{\bar{s}s}/\vev{\bar{q}q}$ and $\chi_{5} \equiv \vev{\bar{s}g 
\sigma \cdot G s} / \vev{\bar{q}g \sigma \cdot Gq}$ for the flavor 
SU(3) symmetry breaking.  The gluon condensate is fixed to $\vev{{ 
\alpha_{s} \over \pi} GG} = (0.36 \GeV)^{4}$.  The vacuum saturation is assumed for 
evaluating the matrix element of the four-quark operators, i.e.\ 
$\vev{(\bar{q}q)^{2}} = \vev{\bar{q}q}^{2}$.  These parameters have some uncertainty, which depends on  the 
truncation in the operator product expansion.  In order to remove this uncertainty, we use  our sum rules in the following way.  The value of $\langle \bar{q} q \rangle$ and $m_{0}^{\;2} 
\equiv \vev{\bar{q} g \sigma \cdot G q}/ \vev{\bar{q}q}$ are 
determined so that the sum rules (\ref{eq:srp}) and (\ref{eq:srn}) 
for the nucleon reproduce the observed masses of $N_{+}$ and $N_{-}$.  
In doing so we require that the prediction of the sum rule at $M 
\simeq m_B$ coincides with the observed mass within 5\% and also that 
for the Borel stability variation of the predicted mass against $M$ in 
the region $m_{B} \sim m_{B}+0.5 \GeV$ is less than 10\%.  In the same 
way, the values of $m_{s}$, $\chi$ and $\chi_{5}$ are determined so 
that the sum rules (\ref{eq:srp}) for the positive-parity hyperons give 
the observed masses of the $\Lambda_{+}$, $\Sigma_{+}$ and $\Xi_{+}$.

\begin{table}[b]
  \caption{The determined QCD parameters}
  \begin{center}
  \begin{tabular}{ccccc}
	\hline \hline
	$\langle\bar{q}q\rangle$ & $m_{0}$ & $m_{s}$ & $\chi$ & $\chi_{5}$  \\

	(-0.244 GeV$)^{3}$ & 0.9 GeV & 0.1 GeV & 0.75 & 0.8   \\
	\hline \hline
	\end{tabular}
  \end{center}
 	\protect\label{tb:detpar}
\end{table}

The determined parameters
are given in Table \ref{tb:detpar}\cite{jo}.
The masses of the positive-parity hyperons, $\Lambda_+$, $\Sigma_+$
and $\Xi_+$ are sensitive to the ${\rm SU}_{f}(3)$ breaking parameters 
$m_s$, $\chi$ and $\chi_5$. Therefore these parameters are determined
well. We see that $\vev{\bar{q}q}$ determines the mass splitting of $B_+$ and $B_-$.

The masses of the flavor octet and singlet baryons calculated in the 
QCD sum rule are shown in Table \ref{tb:result}.  The observed masses 
are reproduced fairly well.  The masses of $\Lambda_-$, 
$\Sigma_-$, $\Xi_-$, $\Lambda_S-$ and $\Lambda_S+$ are the prediction 
without adjustable parameters.  While these masses are 
evaluated at the Borel 
mass $M \simeq m_B$, they are  almost stable against 
the Borel mass, that indicates that the sum rules work well in these 
cases.   The excited $\Xi$ baryon with $J^{P}= {1 \over 
2}^{-}$ has not been identified by experiment\cite{PDG}, but resonances with 
unknown spin and parity are found at 1690 MeV and 1950 MeV. 
The sum rule prefers $\Xi(1690)$.  

Our result suggests 
that the  $B_-$ masses tend to be degenerate.  This is the 
result of two different origins of the mass difference.  The strange 
quark mass raises the hyperon masses, while
the quark condensate widens the mass splitting of $B_+$ and $B_-$.
Because the strange quark condensate is smaller than the 
up and down quark condensate, the effect of the strange quark mass is partly
canceled in the negative parity baryons.

\begin{table}[t]
\begin{center}
  \caption{}
  \begin{tabular}{r|cccccccc|cc}
    \multicolumn{11}{r}{Unit: GeV} \\
	 \hline	\hline
     & $N_+$ &  $\Lambda_+$ & $\Sigma_+$ & $\Xi_+$ & 
            $N_-$ & $\Lambda_-$ & $\Sigma_-$ & $\Xi_-$ 
          & $\Lambda_{S-}$ & $\Lambda_{S+}$	\\
        Cal. & 0.94 & 1.12 & 1.21 & 1.32 & 
	           1.54 & 1.55 & 1.63 & 1.63 & 1.31 & 2.94 \\
	Exp. & 0.94 & 1.12 &1.19  & 1.32 &
	          1.535 & 1.67 & 1.62 & -----& 1.405 & ----- \\
	\hline \hline            
  \end{tabular}
	  \label{tb:result}
\end{center}
\end{table}

It is extremely interesting to observe that the QCD sum rule predicts 
the flavor-singlet $\Lambda_{S}$ spectrum in the reversed order.  
Namely, the baryon $\Lambda_{S-}$ is lighter than the positive parity 
$\Lambda_{S+}$.  This is consistent with the quark model prediction 
that the Pauli principle forbids all quarks occupying the ground 
s-wave state.  In the correlation function $B$ for the $\Lambda_S$, 
there is no dimension five term, $\langle \bar{q} g \sigma \cdot G q 
\rangle$.  If we put the dimension five term in the $B$ correlation 
function by hand and calculate the masses of the $\Lambda_S+$ and 
$\Lambda_S-$, then we find that the increase of the dimension five 
term raises the mass of $\Lambda_S-$ and lowers that of $\Lambda_S+$.  
Thus we conclude that the absence of the mixed condensate term in the 
$\Lambda_{S}$ sum rule causes the reversed order of $\Lambda_{S+}$ and 
$\Lambda_{S-}$.  We also confirm that the $\vev{\bar{q}g \sigma \cdot 
G q}$ terms, is essential in raising the $B_{-}$ masses as was 
stressed in Ref.\cite{lk}.

We note several other approaches of the QCD sum rule for $B_-$.  In 
the case that $B_-$ is the lowest energy state in the considering 
spectrum, its mass is extracted in the usual sum rule.  In 
Ref.\cite{i}, Ioffe pointed out that the negative-parity resonance is 
the ground state in the spectrum of the baryon with spin $J={3 \over 
2}$ and isospin $T={1 \over 2}$ and calculated its mass.  Liu also 
applied the QCD sum rule to $\Lambda (1405)$ and concluded 
that for the $\Lambda (1405)$ the IF consisting of three quarks and a 
flavor octet quark-antiquark pair is important\cite{liu}.  In his analysis, 
however, the continuum term is not considered.  We here obtain the 
realistic mass of $\Lambda(1405)$ in the three-quark sum rule with a 
continuum term.
Some other approaches treat $B_-$ with an IF which does not couple to 
the positive-parity baryon.  In Ref.\cite{cdks} an 
optimized IF for $N_{-}$ was proposed by requiring that the chiral odd 
correlation function, in which the $B_+$ contribution and the $B_-$ 
contribution have different sign, becomes negative.  Lee and Kim\cite{lk}  also 
investigated the mass of N(1535) and 
$\Lambda(1405)$ employing a 
new IF with a covariant derivative, expecting that it has a large 
overlap with the nonrelativistic quark wave function of N(1535).  They 
chose the IF so that it does not couple to the ground state 
nucleon\cite{lk}.  We, however, employ the nonderivative IF in 
the present study because our main interest is to study the mechanism 
of $B_{+}$--$B_{-}$ mass splitting and difference in their properties.

\section{Meson Couplings to the Negative Parity Baryons}

Concerning the meson-baryon couplings, the negative parity baryons,
$N(1535)$, $\Lambda(1670)$ and $\Sigma(1750)$
(denoted by $B^{*}$), possess interesting
properties.
The most distinguished feature is
their relatively large decay widths of $B^* \to \eta B$\cite{PDG}.
Because of the smallness of the available phase space,
this fact suggests relatively
large coupling constants of $\eta BB^*$ as compared to those of
$\pi BB^*$.

One may also look at the problem in the following way.
Using the experimental decay widths of the resonances, we obtain, for
example,
$\gpNR \sim 1$ and $\geNR \sim 2$.
These values are in fact much smaller than those in the $NN$ sector:
$\gpNN \sim 13$ and $\geNN \leq 5$.
Furthermore, the pion couples weaker than the eta in the $NN^*$ sector,
as
opposed to the $NN$ sector.
Thus, one may ask why the coupling $\gpNR$ is suppressed so much
as compared with other couplings\cite{joh}.

In order to apply the QCD sum rule to the meson-baryon vertices, 
we follow the method
used by Shiomi and Hatsuda\cite{sh}.
They studied the $\pi NN$ coupling constant $\gpNN$
by using the two point function between
the vacuum and a one meson state in the soft meson limit $(q^{\mu}
\to 0)$.
The relevant correlation function is
\beq
   \Pi^{m}(p) &=& i \int d^4 x \, e^{ip\cdot x}
        \bra{0} TJ_{N}(x) \bar{J_{N}}(0) \cket{m(q=0)} \nonumber \\
        &=&
        i \gamma_{5} (\Pi_{0}^{m}(p^{2})  + \Pi_{1}^{m}(p^{2}) \pslash ) \, ,
   \label{eq:cor}
\eeq
where $J_{N}$ is defined in (\ref{eq:nucur}), and $m$ denotes
either $\pi$ or $\eta$.
The parameter $t$ will be chosen suitably depending on whether $J_{N}$
should be coupled strongly to positive or negative parity baryons.

Let us first look at the phenomenological side of the $\pi-N-N^*$ correlation
function to see how this coupling strength can be extracted.
We define the phenomenological $\pi NN^*$
interaction Lagrangian,
\begin{equation}
    \label{eq:defpNN}
    {\cal L}_{\pi NN^*} = \gpNR \bar{N}^{*} \tau^{i} \pi^{i} N,
\end{equation}
where $N$ and $N^{*}$ are the field operators for the positive
and negative parity nucleons,
$\pi^{i}$ is the pion field, and $\tau^{i}(i =
1,2,3)$ are the Pauli matrices for isospin.
From the
Lagrangian (\ref{eq:defpNN}), the $\pi NN^{*}$ contribution in the
$\Pi^{\pi}(p)$ is given in the soft pion limit by
\begin{equation}
   \label{eq:pNN*}
    \gpNR \lambda_{N} \lambda_{N^{*}}
    \left[{p^{2} + m_{N}m_{N^{*}} \over (p^{2} - m_{N}^{2}) (p^{2} -
    m_{N^{*}}^2)} + { \pslash (m_{N} + m_{N^{*}}) \over (p^{2} -
    m_{N}^{2}) (p^{2} - m_{N^{*}}^2)}\right]i\gamma_{5} \,
\end{equation}
where $\lambda_{N}$ and $\lambda_{N^{*}}$ are defined by
$\langle 0 | J_{N} | N\rangle = \lambda_{N} u_{N}$ and $\bra{0} J_{N}
\cket{N^{*}}= \lambda_{N^*} i\gamma_5 u_{N^*}$, respectively, with
$u_{B}$ being the Dirac spinor for the baryon $B$.
We note that  there appear two terms in (\ref{eq:pNN*}); one
proportional to $\gamma_{5}$ and the other proportional to
$\pslash \gamma_{5}$.
In contrast, the $\pi NN$ contribution has only one term,
\begin{equation}
   \gpNN \lambda_{N}^{\;2} {i \gamma_{5} \over
    p^{2} - m_{N}^{\; 2}} \, ,    \label{eq:pheNN}
\end{equation}
as is derived from the $\pi NN$ interaction Lagrangian
\begin{equation}
    {\cal L}_{\pi NN} = \gpNN \bar{N} i \gamma_{5} \tau^{i} \pi^{i}
    N\, .
\end{equation}
We note that (\ref{eq:pheNN}) is also obtained by replacing
$M_{N^*}$ by $-M_N$ in (\ref{eq:pNN*}).

In the soft pion limit,
Shiomi and Hatsuda\cite{sh} studied the sum rule using the non-vanishing term
of (\ref{eq:pheNN}) and found that the resulting $\pi NN$ coupling constant
satisfies the
Goldberger-Treiman relation with $g_{A}=1$.
Recently, Birse and Krippa also studied the coupling constant
$g_{\pi NN}$ at a non zero pion momentum\cite{bk}.
For the $\pi NN^*$ coupling constant
we study the term proportional to $\pslash \gamma_{5}$,
which is expected to have
a dominant contribution from the $\pi NN^*$ coupling.
In fact, there could be a contribution in the $i\pslash \gamma_{5}$
term from positive parity resonances;
a dominant part would be from the lowest resonance $N(1440)$.
Such a term is, however, proportional to
the mass difference $M_{N(1440)} - M_{N}$ unlike the sum as in
the second term of (\ref{eq:pNN*}).
Thus the contribution from $N(1440)$ will be relatively
suppressed as compared with that of $N(1535)$.
Moreover since we choose the
mixing parameter $t \sim 0.8$ such that
the interpolating field (\ref{eq:nucur})
couples strongly to negative parity states, we expect least
contamination from positive parity resonances.

The sum rule for the $\eta NN^{*}$ coupling is
similarly constructed
by replacing the isospin matrices $\tau$ in the $\pi NN^*$
coupling by the unit matrix.

The correlation function is now computed by the operator product
expansion (OPE) perturbatively in the deep Euclidian region.
The result for the terms of $i \pslash \gamma_{5}$ takes the
following form
\begin{eqnarray}
    \Pi^{\rm OPE}(p) & = & i \int d^{4}x \, e^{ip\cdot x} \, \bra{0}{\rm T}
      J_{N}(x;s) \bar{J}_{N}(0;t) \cket{m}
      \label{eq:PiOPE}\\
    & \equiv & i \pslash \gamma_{5} \left[ {\cal C}_{4} \ln(-p^{2}) +
        {\cal C}_{6} {1 \over p^{2}} 
       + \cdots
        \right] + i\gamma_{5} \left[ {\cal C}_{3} p^{2} \ln(-p^{2}) + \cdots
        \right], \nonumber
\end{eqnarray}
where we allow to use the different mixing parameters in the interpolating
fields such that $\bar{J}_N (0; t \sim 0.8)$ couples dominantly to the
$N^*$ state,
while $J_N(x; s = -1)$ to the $N$ state.
Note that the terms of $i \pslash \gamma_{5}$ are of even dimension.
The correlation function (\ref{eq:PiOPE}) has been calculated up to
dimension 8, ignoring higher order terms in $m_{q}$ and $\alpha_{s}$.
The results are
\begin{eqnarray}
{\cal C}_{4} & \sim &
        m_{q} \bra{0} \bar{q}i \gamma_{5} q \cket{m}
        \stackrel{m_q \to 0}{\longrightarrow} 0 \\
{\cal C}_{6} & = & -{s-t \over 4} \left[
    \langle \bar{d}d \rangle \bra{0} \bar{u} i \gamma_{5} u \cket{m} +
    \langle \bar{u}u \rangle \bra{0} \bar{d} i \gamma_{5} d \cket{m}
     \right] \\
{\cal C}_{8} & = & - {s-t \over 144} \left[
    25 (\langle \bar{d}gG \cdot \sigma d \rangle
        \bra{0} \bar{u} i \gamma_{5} u \cket{m} +
    \langle \bar{u}gG \cdot \sigma u \rangle
        \bra{0} \bar{d} i \gamma_{5} d \cket{m}) \right.  \nonumber \\
    & &  \left. - 7 ( \langle \bar{d}d \rangle
        \bra{0} \bar{u} i \gamma_{5} gG \cdot \sigma u \cket{m} +
     \langle \bar{u}u \rangle
        \bra{0} \bar{d} i \gamma_{5} gG \cdot \sigma d \cket{m}) \right]
\end{eqnarray}
where
we assume the vacuum saturation for four-quark matrix elements.
We evaluate the meson-vacuum matrix elements using 
the soft meson theorem, and obtain the following relations:
\begin{eqnarray}
  \label{eq:uum}
    \bra{0} \bar{u}i \gamma_{5} u \cket{m} & = & -
        {\alpha_{m} \over f_{m}} \langle \bar{u}u \rangle \, , \\
  \label{eq:ddm}
    \bra{0} \bar{d}i \gamma_{5} d \cket{m} & = & \pm
        {\alpha_{m} \over f_{m}}  \langle \bar{d}d  \rangle \, ,  \\
  \label{eq:uGum}
    \bra{0} \bar{u}i \gamma_{5} G\cdot \sigma u\cket{m}
     & = & - {\alpha_{m} \over f_{m}}
         \langle \bar{u}G\cdot \sigma u \rangle  \, , \\
  \label{eq:dGdm}
    \bra{0} \bar{d}i \gamma_{5} G\cdot \sigma d \cket{m}
     & = & \pm {\alpha_{m} \over f_{m}}
         \langle \bar{d} G\cdot \sigma d \rangle \, ,
\end{eqnarray}
where $\alpha_{\pi} = 1/\sqrt{2}$ and $\alpha_{\eta} = 1/\sqrt{6}$.
Note that the sign  difference between
the pion and eta matrix elements comes from the isospin structures:
$\pi^{0} \sim \frac{1}{\sqrt{2}}(\bar{u}u - \bar{d}d)$,
and $\eta \sim \eta_8
\sim \frac{1}{\sqrt{6}}(\bar{u}u + \bar{d}d - 2 \bar{s}s)$
(by neglecting small mixing angle effects).
We note that the $\bar s s$ component in $\eta$ is irrelevant up to
dimension 8, since the
interpolating field (\ref{eq:nucur}) does not contain
strange quarks.

From (\ref{eq:PiOPE}) -- (\ref{eq:dGdm}),
we find that the correlation function for the $\pi NN^{*}$ coupling
vanishes in the chiral limit $m_q \to 0$, and therefore $\gpNR=0$.
In contrast,  the correlation function for the $\eta NN^*$
coupling does not vanish, and so
the coupling constant $\geNR$ remains finite.
We emphasize that the result from the OPE here
is qualitatively consistent with the partial decay rates of $N(1535)$,
$\Gamma_{N(1535) \to \pi N} \sim \Gamma_{N(1535) \to \eta N}
\sim 70$ MeV.

The vanishing correlation function for the $\pi NN$ case is, in
fact, a general consequence of chiral symmetry.
We might have applied the soft meson theorem to the correlation function
(\ref{eq:cor}) from the beginning.
Using the transformation property
$[ Q_5^a , J_N] = i \gamma_5 \tau^a J_N$, we find
\begin{eqnarray}
\Pi^{\pi^a} (p) &=& \lim_{q \to 0}
\int d^4x e^{ipx} \bra{0} T J_N(x) \bar J_N(0) \ket{\pi^a(q)}
\nonumber \\
&=& - \frac{i}{\sqrt{2}f_\pi} \int d^4x e^{ipx} \bra{0} [ Q_5^a ,
T J_N(x) \bar J_N(0) ] \ket{0} \nonumber \\
&=&  \frac{1}{\sqrt{2}f_\pi}
\int d^4x e^{ipx}
\{ \gamma_5\, \tau^a , \bra{0} T J_N(x) \bar J_N(0)  \ket{0} \} \, .
\label{Q5comm}
\end{eqnarray}
In the last expression, we have recovered the vacuum to vacuum
transition,
which has a Lorentz structure
$\bra{0}  J_N(x) \bar J_N(0)  \ket{0} \sim A \pslash + B 1$.
Thus in (\ref{Q5comm}), the term of $\pslash \gamma_5$ disappears.
We emphasize that this is a consequence of chiral symmetry.
In order to achieve this result, the nucleon current has to
transform appropriately under chiral transformations and also, the
negative 
parity baryons are produced by the same nucleon current $J_N$.
These two facts determine the chiral properties of the positive and
negative parity nucleons.
As a consequence, the $\pslash \gamma_5$ term, which in
general should exist in the two point function, is shown to disappear.

What are then implications of the vanishing matrix element of
$\pslash \gamma_5$ term?
By relating this with the sum rule, it is implied that the coherent
sum over all coupling constants of the pion between various baryon
excitations will vanish.
One would, in fact,
make a stronger statement by using properties of analytic functions.
Namely, if an analytic function vanishes in a certain domain, it
vanishes
identically for whole analytic region.
In the spirit of the QCD sum rule, we have found that the
$\pslash \gamma_5$ term of the correlation
function vanishes in the deep Euclidian region, which implies the
identically vanishing $\pslash \gamma_5$ term for whole momentum space.
There might be an exceptional case, if there are two
(delta-function like) terms having equal strengths but
with opposite signs, which sum up to be zero.
In fact, the low-lying negative parity nucleons do look like that;
there are two neighboring states of $N(1535)$ and $N(1650)$.
Even in this case, one would say that
a state which is coupled by the current $J_N$ is a
superposition of two (or more) degenerate states,
and couplings with the pion is
coherently added up and cancel.

One may wonder if such a suppression of $\gpNR$
could be explained in some way by spontaneously broken chiral symmetry.
From chiral symmetry point of view, it seems natural to put $N$ and
$N^*$
in the same multiplet of chiral partner (or parity doublet).
There have been several attempts to treat positive and negative
parity baryons in this point of view.

DeTar and Kunihiro considered the parity doublet nucleons in the linear
sigma model of $SU(2) \times SU(2)$\cite{dk}.
In addition to the standard chiral invariant interaction terms,
they introduced a chiral invariant mass term
between the positive and negative parity baryons.
The strength $m_{0}$ for the non-standard mass term
reduces to the mass of the would-be chiral doublet nucleons when the
chiral symmetry restores.
In the spontaneously broken phase,
the mass splitting is proportional to the non-zero
value of the quark condensate.
In this model, it has been shown that
$\gpNR$ is proportional to $m_{0}$  to the leading order
in $m_{0}$.
Therefore, if $m_0=0$, the coupling constant $\gpNR$ vanishes.
This is a rigorous consequence from chiral symmetry.
The question, is therefore, whether the non-standard mass term exists or
not in the real world.
This can be examined by looking at baryon masses in the chiral symmetric
phase.
In the QCD sum rule study\cite{jko}, the masses of $N$
and $N^*$ seem to be degenerate and decrease as the quark condensate
$\langle \bar q q \rangle$ is decreased.
This implies a small (and possibly vanishing) $m_{0}\approx 0$
and so a small
$\gpNR$.

The formulation of DeTar and Kunihiro may be extended to the chiral U(1)
$\times$ SU(2) model, where the eta is a unitary singlet.
In this extension, there seems to exist a natural explanation for
the relation among the coupling constants:
$\gpNN \gg \geNN$, while $\gpNR \ll \geNR$.

\section{Conclusion}

We have applied the QCD sum rule method to the study of the negative-
parity baryons.  The masses of the flavor octet and singlet baryon 
resonances are reproduced fairly well. An interesting observation is near 
degeneracy of the octet members, that seems to come from the interference 
between two different mechanisms of the symmetry breaking.  The flavor 
singlet $\Lambda_S$ has the negative-parity state as the lowest mass, that is 
consistent with experiment.  The formulation and the numerical results 
suggest a picture that the positive and negative-parity baryons form a parity 
doublet, which would be degenerate when the chiral symmetry is
restored.

The calculation of the $\pi NN^*$ coupling has been performed and gives a 
new surprise, that  the coupling vanishes in the chiral limit.  
We suggest that 
the chiral symmetry is responsible for this result.  It seems consistent with 
the observation of suppressed pion decay rate of $N(1535)$.  We suggest 
that these properties of $N^*$ can be understood in a chiral effective theory 
for the parity doublet nucleons proposed by Detar and Kunihiro\cite{dk}.

\bigskip

The authors would like to thank the organizers of the INT/CEBAF $N^*$ 
Workshop for giving them chances to participate the exciting workshop 
and also acknowledge INT and CEBAF for their support.

\end{document}